\documentclass[useAMS,usenatbib]{mn2e}
\usepackage{amsmath,amssymb,mathrsfs,graphicx,float,latexsym}
\usepackage{epstopdf,ragged2e}
\usepackage{hyperref}

\newcommand{\ep}{\epsilon_{p}}

\newcommand{\tsd}{\tau_{\text{sd}}}

\def\apj{{ApJ}}

\def\apjl{{ApJL}}

\def\04a{{2004 a}}
\def\04b{{2004 b}}

\title[Precessing magnetars as central engines in SGRBs]{Precessing magnetars as central engines in short gamma-ray bursts}
\author[A. G.~Suvorov and K. D.~Kokkotas]{A. G.~Suvorov\thanks{E-mail: arthur.suvorov@tat.uni-tuebingen.de} and K. D.~Kokkotas\\Theoretical Astrophysics, Eberhard Karls University of T{\"u}bingen, T{\"u}bingen, D-72076, Germany}
\begin{document}

\date{Accepted ?. Received ?; in original form ?}

\pagerange{\pageref{firstpage}--\pageref{lastpage}} \pubyear{?}

\maketitle
\label{firstpage}

\begin{abstract}

\noindent{Short gamma-ray bursts that are followed by long-duration X-ray plateaus may be powered by the birth, and hydrodynamic evolution, of magnetars from compact binary coalescence events. If the rotation and magnetic axes of the system are not orthogonal to each other, the star will undergo free precession, leading to fluctuations in the luminosity of the source. In some cases, precession-induced modulations in the spin-down power may be discernible in the X-ray flux of the plateau. In this work, 25 X-ray light curves associated with bursts exhibiting a plateau are fitted to luminosity profiles appropriate for precessing, oblique rotators. Based on the Akaike Information Criterion, 16 (64 per cent) of the magnetars within the sample display either moderate or strong evidence for precession. Additionally, since the precession period of the star is directly tied to its quadrupolar ellipticity, the fits allow for an independent measure of the extent to which the star is deformed by internal stresses. Assuming these deformations arise due to a mixed poloidal-toroidal magnetic field, we find that the distribution of magnetic-energy ratios is bimodal, with data points clustering around energetically equal and toroidally dominated partitions. Implications of this result for gravitational-wave emission and dynamo activity in newborn magnetars are discussed.
}

\end{abstract}


\section{Introduction}

Short gamma-ray bursts (SGRBs) are likely associated with compact binary coalescence events involving at least one neutron star (NS), as evidenced by the dual detection of GW170817 and GRB 170817A \citep{abb17,gold17}. For NSNS binaries where the progenitor stars are not too massive, the merger may give birth to a millisecond (proto-)magnetar rather than a black hole, which subsequently helps to fuel the burst through the launching of a magnetic wind \citep{usov92,dai98,met08,bucc11,piro17}. While some studies suggest that the remnant must be hypermassive and collapse within $\lesssim 100$ ms after formation, so that relativistic outflows are not choked by matter ablated from the stellar surface via neutrino fluxes \citep{thom04,lee07,mur14}, many sources exhibit late-time X-ray flares \citep{camp06,mar11} and afterglow emissions, lasting up to $\sim 10^{5}$ s in some cases \citep{fan06,tang19} [see \cite{ciolfi18} and \cite{sarin20b} {for recent reviews}]. Extended emissions of this sort suggest persistent energy injections from a long-lived central engine, consistent with numerical simulations finding that stable magnetars may form from NSNS mergers \citep{giac13,ciolfi17,radice18,mosta20}.

Afterglows that are characterised by a roughly constant `plateau' phase in the X-ray flux followed by a power-law fall-off are especially suggestive of a magnetar central engine \citep{zhang01,met11,lasky16}. In this case, angular momentum is carried away from the system by the magnetar wind, which accelerates the surrounding ejecta and drives a forward shock through the ambient medium \citep{gao13,gao15}. In particular, the radiation energy from such a source is expected to be roughly constant until a spindown time has elapsed, after which the luminosity associated with the progressively slower remnant decays quadratically. On the other hand, several SGRBs exhibit light curves with sharp fall-offs in the flux $\sim 10^{2}$ s after the prompt emission \citep{tang19}. Systems displaying these `internal plateaus' may instead contain a NS in a supramassive state \citep{rowl10,lasky14}, where the baryonic mass exceeds the hydrostatic limit though the star resists collapse thanks to centrifugal forces \citep{breu18,rezz18}. Once the object decelerates sufficiently, collapse sets in and the signal is truncated. In any case, detailed properties of the remnant may be inferred from the character of the X-ray afterglow \citep{usov92,rowl13,lu15,stratta18,sarin20}. 

The electromagnetic spindown luminosity is directly proportional to a trigonometric function of the inclination angle, $\alpha$, made between the magnetic and rotation axes of the star  \citep{spit06,kara09,phil15}. Turbulent convection in the post-merger remnant is expected to erase any pre-existing correlation that may exist between the Euler angles of the system \citep{thom93,melatos00,lander18}, and thus it is unlikely that the magnetar acts as an orthogonal rotator at birth, independently of environmental details. The star will thus be an oblique rotator in general, whose spindown luminosity fluctuates as the inclination angle `wobbles' due to magnetic field and rotational profile evolution. At least in the early stages of the star's life, either free or torque-driven precession may be the dominant driving factor for the evolution of $\alpha$ \citep{gold70,lai15} [though cf. \cite{hask20}]. Precession periods on the order of seconds, as expected for a rapidly rotating and highly elliptical remnant, coincide with the periodicities of gamma-ray flux oscillations observed in several GRBs \citep{hard91,usov92,marg08}. At least in the case of GRB 090510, \cite{suv20} recently showed that there is a statistical preference for a precessing model over an orthogonal model in describing afterglow data. 

Aside from oscillations in X-ray afterglows, precession has been invoked to explain a number of neutron-star phenomena. Pulsar timing has revealed correlated variations in the pulse shape and spindown rate of the isolated object PSR B1828-11 \citep{stairs00}, suggesting a precessing source \citep{link01}. The consistent increase in pulse separations \citep{lyn13} together with measurements of a braking index $n<3$ \citep{lyn98} in the Crab pulsar are also suggestive of precessional evolution in the inclination angle \citep{lai15}. Sinusoidal variations in the thermal spectrum of 1E 61348-5055 \citep{heyl02} and in the hard X-ray spectra of 4U 0142$+$61 \citep{makis14} and Her X-1 \citep{post13} provide further evidence for free precession in NSs. Recently, \cite{amiri20} reported the detection of a $16.35 \pm 0.15$ day periodicity in the repeat-burster FRB 180916.J0158+65, which has been interpreted as the free \citep{lev20,zan20} or torque-induced \citep{sob20} precession period of the source. Given the long-hypothesied connection between repeating fast radio bursts and magnetar activity \citep{lyu14,belo17,suvk19}, and the recent establishment of the galactic magnetar SGR 1935$+$2154 as a source of such bursts \citep{boch20}, the object responsible for FRB 180916.J0158+65 may indeed be a precessing magnetar. 

It is the purpose of this paper to systematically explore the role of magnetar precession in GRB afterglows. To this end, we extend the previous study of \cite{suv20}, who fitted orthogonal and precessing models to GRB 090510, to include a much larger sample (25) of SGRBs. The sample we use is a subset (see Sec. 3) of those sources identified by \cite{rowl13} and \cite{lu15} which exhibit the plateau phase expected of a magnetar central engine. Although a multitude of effects can influence spindown beyond the simple orthogonal versus precessing possibilities considered here (see Sec. 2.1), this study provides, for the first time, a statistical look at the genericity of precession in newborn NSs.

This paper is organised as follows. In Section 2 we briefly review the magnetar central engine model for SGRBs, and derive expressions relevant for the dipolar, spindown luminosity of both orthogonal and precessing rotators. Section 3 covers the light-curve fitting methods and details the sample of SGRBs used throughout. In Section 4, the fitted parameters are given together with an analysis of their statistical significance. Some discussion is given in Section 5. A concordance cosmology with \cite{planck18} parameters $H_{0} = 67.4 \text{ km s}^{-1} \text{ Mpc}^{-1}$, $\Omega_{M} = 0.32$, and $\Omega_{\Lambda} = 0.68$ is adopted throughout to translate between observed fluxes and (bolometric) source luminosities. 

\section{Magnetar central engine for short gamma-ray bursts}

In the magnetar central-engine model for GRBs, extended X-ray emissions may be powered by the pumping of energy into the forward shock which propagates through the ambient medium, via the conversion of mechanical energy into radiation energy \citep{gao13,gao15}. In particular, the energy reservoir of the newborn star is primarily composed of its rotational kinetic energy\footnote{Throughout this work, we adopt the convention $Z_{x} = 10^{-x} Z$ for quantities $Z$ in CGS units. Mass is an exception, for which we use $M_{x} = \left(x \times M_{\odot}\right)^{-1} M$.} \citep{usov92},
\begin{equation} \label{eq:energy}
E_{\text{rot}} = \frac {1} {2} I_{0} \Omega^2 \approx 3.1 \times 10^{52} M_{2.0} R_{6}^{2} P_{-3}^{-2} \text{ erg},
\end{equation}
where $I_{0}/(M_{2.0} R_{6}^2) \approx 1.6 \times 10^{45} \text{ g cm}^2 $ is the moment of inertia for stellar radius $R$ and mass $M$, and $P$ denotes the spin period of the star. Spindown, driven by electromagnetic torques or otherwise [e.g., gravitational radiation \citep{don15}], ultimately converts the kinetic energy \eqref{eq:energy} into radiation energy with luminosity $L$, viz.
\begin{equation} \label{eq:spindownrot}
-\dot{E}_{\text{rot}} \approx - I_{0} \Omega \dot{\Omega} = L,
\end{equation}
where the approximation symbol is used because we ignore effects that lead to a time-dependent moment of inertia, such as mass ablations driven by neutrino outfluxes, which are expected to be of the order $\dot{M} \lesssim 5 \times 10^{-4} M_{\odot} \text{ s}^{-1}$ \citep{thom04,lee07,dess09}, and fallback accretion, which could potentially occur at super-Eddington rates shortly after birth with $\dot{M}_{a} \lesssim 0.1 M_{\odot} \text{ s}^{-1}$ \citep{metz18}.

In writing down expression \eqref{eq:spindownrot}, we have additionally assumed an idealised conversion process. In reality, the X-ray radiation that is actually observed results from complicated interactions between a photon-pair plasma nebula, inflated by the spindown radiation, and the optically-thick material surrounding the remnant that was ejected during the main phase \citep{ciolfi18}. The energy reprocessing that powers the plateau is therefore not perfectly efficient. An efficiency parameter $\eta \leq 1$, which is likely to be time-dependent in reality \citep{xiao19}, can be coupled to $L$ to account for this. Although we effectively set $\eta$ to unity throughout, the scaling with $\eta$ is given when introducing fitted values.

\subsection{Electromagnetic dipole radiation}

For simplicity, and to make a quantitative statistical comparison with the results of \cite{rowl13} [see also \cite{met11}], we assume that the newborn NS spins down only due to electromagnetic torques associated with a centered dipole. In general, higher-order multipoles \citep{krol91,pet19,suvm20}, fallback accretion from gravitationally bound debris \citep{ross07,mel14,gib17}, stellar mass losses from neutrino outflows \citep{thom04,dess09} or disk mass losses from ${}^{4}\text{He}$ synthesis via free nuclei recombinations \citep{lee09,mur14}, and gravitational radiation due to magnetic deformations \citep{ciolfi09}, quasi-normal oscillations \citep{kru20}, or precession itself \citep{gao20}, all of which may help to facilitate spin-flip \citep{dall18} or bar-mode \citep{don15} instabilities, affect the luminosity of the central engine. Inclusion of these effects leads to a system with many free parameters, which obfuscates the potential role of the magnetic inclination angle and becomes difficult to constrain. We also ignore contributions from fireball shocks \citep{sarin19}, which may work in conjunction with the continuous injection from the magnetar to power the extended emission. 

The electromagnetic spindown luminosity of a dipole with polar field strength $B_{p}$ is given by [e.g., \cite{lu15}]
\begin{equation} \label{eq:spindown}
L_{\text{EM}} =  \frac {B_{p}^2 R^{6} \Omega^4} {6 c^3} \lambda(\alpha),
\end{equation}
for polar field strength $B_{p}$, where $\lambda$ is a magnetospheric factor that depends on the orientation of the NS, primarily through the inclination angle $\alpha$ \citep{kara09,phil15}. In a pure vacuum, one has well-known result $\lambda(\alpha) = \sin^{2}\alpha$, and the spindown power is maximal for an orthogonal rotator with $\alpha = \pi/2$. However, newborn stars are generally expected to be surrounded by magnetised plasma \citep{gj69}, especially just after birth \citep{uz06,ciolfi18}. In any case, \cite{spit06} found that $\lambda(\alpha) \approx k_{1} + k_{2} \sin^2 \alpha$ with $k_{1} = 1 \pm 0.05$ and $k_{2} = 1 \pm 0.11$ provided an excellent fit to numerical simulations of force-free magnetospheres. To a good approximation, it is therefore reasonable to take $\lambda(\alpha) = 1 + \sin^2 \alpha$. In the previous work of \cite{suv20}, an ignorance factor $\delta$ was included through $\lambda(\alpha) = 1+ \delta \sin^2 \alpha$, which leads to a more diverse range of magnetospheric environments. However, it is unclear whether it is physical to allow $\delta$ to vary substantially between different GRBs, so we set it to unity here to reduce the dimensionality of the parameter space.

\subsection{Precessing, oblique rotators}

At early times $t \lesssim \tsd$ relative to birth, where $\tsd \sim 3.3 \times 10^{3} \text{ s} \times \left( M_{2.0} B_{p,15}^{-2} P_{0,-3}^{2} R_{6}^{-4} \right) $ is the characteristic spindown time for birth spin-period $P_{0} = 2 \pi / \Omega_{0}$, the inclination angle of the star evolves mainly due to precession \citep{gold70,lai15} [though cf. \cite{hask20}], viz.
\begin{equation} \label{eq:alphaeqn}
\dot{\alpha} \approx k \Omega_{p}  \csc \alpha \sin(t \Omega_{p} ),
\end{equation}
where $k$ is an order-unity factor related to the Euler angles $\chi$ and $\theta$ [see equation (50) in \cite{lai15}], and 
\begin{equation} \label{eq:precfreq}
\Omega_{p} = \epsilon_{p} \cos \theta \Omega
\end{equation}
 is the precession frequency for rotationally-misaligned ellipticity\footnote{Note that $\epsilon_{p}$ only accounts for distortions that are not aligned with the rotation axis. In contrast, the rotational ellipticity (sometimes called the zero-strain oblateness) of a newborn NS is expected to be large, $\epsilon_{\text{rot}} \approx 0.15 R_{6}^3 M_{2.0}^{-1} P_{-3}^{-2}$, though this oblateness does not contribute to precession \citep{lai15}.} $\epsilon_{p}$. 

The task of determining the spindown luminosity for a precessing, oblique rotator now becomes one of solving the coupled system of non-linear equations \eqref{eq:spindownrot} and \eqref{eq:alphaeqn} subject to some set of initial conditions. While finding numerical solutions to the coupled system \eqref{eq:spindownrot} and \eqref{eq:alphaeqn} is manageable in principle, it is difficult to subsequently  fit light-curve data; to perform a Monte Carlo search for best-fit parameters (see Sec. 3), a sizeable parameter space must be scanned, and it is computationally infeasible to solve these equations for a large $(\lesssim 10^{8})$ set of initial conditions and nascent parameters. 

Following \cite{suv20} we can, however, obtain an approximate but analytic solution to the coupled system \eqref{eq:spindownrot} and \eqref{eq:alphaeqn} in the following way. In the orthogonal rotator case, we can ignore equation \eqref{eq:alphaeqn} completely since $\dot{\alpha} = 0$. The remaining equation, \eqref{eq:spindownrot}, has exact solution $\Omega(t) = \Omega_{0} \left( 1 + t/ \tsd \right)^{-1/2}$, which, from \eqref{eq:spindown}, reproduces the usual expression for the spindown luminosity of an orthogonal rotator (see below). In principle, if the precession frequency were non-zero, equation \eqref{eq:precfreq} gives us 
\begin{equation} \label{eq:prec0}
\Omega_{p}(t) \approx \epsilon_{p} \Omega_{0} \left( 1 + t/ \tsd \right)^{-1/2} 
\end{equation}
in this case (for $\cos \theta \approx 1$). On the other hand, for $\dot{\Omega} \approx 0$ [valid for $t \lesssim \tsd$ since $\Omega(t \lesssim \tsd)$ is approximately constant so as to produce a plateau] and $\dot{\epsilon}_{p} \approx 0$ (ignoring magnetic field evolution and mass losses), equation \eqref{eq:alphaeqn} can be solved exactly to give 
\begin{equation} \label{eq:alphasln}
\alpha(t) = \cos^{-1} \left[ \alpha_{0} + k \cos \left( t \Omega_{p} \right) \right],
\end{equation}
for some $\alpha_{0}$. We may therefore obtain an approximate though physically reasonable solution to the coupled system \eqref{eq:spindownrot} and \eqref{eq:alphaeqn} by substituting \eqref{eq:prec0} into \eqref{eq:alphasln}. In doing so, solving the energy-balance equation \eqref{eq:spindownrot} yields the precession-modified electromagnetic spindown luminosity
\begin{equation} \label{eq:genspinlum}
\begin{aligned}
L_{p} \approx&  \frac {B_{p}^2 R^6 \Omega_{0}^4} {6 c^3}  \left\{ 2 - \left[ \alpha_{0} + k \cos \left( t \Omega_{p}  \right) \right]^{2} \right\} \\
&\times \Big\{ 1 + \frac {t \left( 2 - \alpha_{0}^2 - \tfrac {1} {2} k^2 \right)} {\tsd} \\
&- \frac {k \left[ 2 \alpha_{0} + \tfrac{1}{2} k \cos \left( t \Omega_{p}  \right) \right] \sin \left( t \Omega_{p} \right)} {\tsd \Omega_{p}} \Big\}^{-2}.
\end{aligned}
\end{equation}

In the orthogonal rotator case, obtained for $k = 0$ and $\alpha_{0} = 1$ [equivalently for $\lambda(\alpha) = \sin^2\alpha$ and $\alpha = \pi/2$], $L_{p}$ reduces to the well-known form
\begin{equation} \label{eq:spiinlum}
L_{\perp} = \frac {B_{p}^2 R^6 \Omega_{0}^4} {6 c^3} \left( 1 + \frac {t} {\tsd} \right)^{-2},
\end{equation}
which is the luminosity profile used by \cite{rowl13} to infer magnetar parameters from a variety of SGRBs.

In this work, we perform two separate fits to afterglow data for each GRB, one using expression \eqref{eq:genspinlum} and another with \eqref{eq:spiinlum}. It is important to emphasise again that the former is only an approximate expression relevant for a precessing, oblique rotator. At times $t \gg \tau_{\text{sd}}$, additional effects modify the evolution of $\alpha$ [see equation (7) in \cite{gold70}], though we neglect such terms to keep the analysis straightforward and analytic. Nevertheless, the results presented here should be considered qualitative in nature, and not representative of a realistic physical process, which requires detailed numerical simulations beyond the scope of this work (see also Sec. 2.1).

\section{Catalogue of magnetar candidates}

By analysing Swift-XRT light curves from short (durations $T_{90} \lesssim 2 \text{ s}$, where $T_{90}$ is defined as the time window in which $90\%$ of the energy is released) GRBs, \cite{rowl13} identified 18 firm and 10 ``possible'' magnetars as central-engine candidates. In that work, candidature was  assessed primarily on the existence of a plateau phase in the flux followed by either a power-law falloff or a steep drop (temporal index $\ll -2$). In particular, from expression \eqref{eq:spiinlum}, we see that a magnetar spinning down to electromagnetic dipole radiation is expected to have a roughly constant luminosity for $t \lesssim \tsd$, while for $t \gtrsim \tsd$ one has instead $L_{\perp} \propto t^{-2}$. By fitting the observed light curves to equation \eqref{eq:spiinlum} (though also accounting for a variety of observational factors; see Sec. 4), \cite{rowl13} estimated the spin periods and polar magnetic field strengths for the newborn magnetars formed in those 28 bursts. Along similar lines, several additional candidates were identified by \cite{lu15}, who also included distinctions between internal and external plateaus, which we do not consider here.

In this work, we perform Monte Carlo simulations aiming to directly minimise the mean-square errors from precessing [equation \eqref{eq:genspinlum}] and orthogonal [equation \eqref{eq:spiinlum}] fits to the BAT-XRT lightcurves of the SGRBs for which magnetar candidates were identified by \cite{rowl13}, together with two from \cite{lu15}. Unfortunately, a handful of candidates from \cite{rowl13} are either seemingly unavailable on the SWIFT catalogue\footnote{\protect\url{http://www.swift.ac.uk/xrt_curves/}} [081024; \citep{evans09}] or have too few $(\lesssim 5)$ data points to perform a meaningful fit (050509B, 080702A, 090621B, 091109B). Subtracting these from the pool of candidates, we eventually analyse 25 light curves. Data pertinent to these GRBs, including the redshift $z$ (in cases where host galaxies were identified) and the photon index $\Gamma_{\gamma}$ as measured by the detector(s), both of which are needed to convert between observed fluxes and source luminosities, are listed in Table \ref{tab:grbpars}.

\begin{table*}
\caption{Observed Properties of the SGRBs used in our sample. All data are taken from (Evans et al. 2009), Rowlinson et al. (2013), L{\"u} et al. (2015), and references therein. For convenience, we also list the luminosity distances $D_{\text{L}}$ of the sources, which are calculated from the listed redshifts using a concordance cosmology with Planck Collaboration et al. (2020) parameters. When the redshift of the source is unknown, a canonical value of 0.72 is used, as in Rowlinson et al. (2013). Light curve data from these sources, used in analysis throughout this work, comes from the Swift-XRT Catalogue (Evans et al. 2009). Error bars are quoted at the $90\%$ confidence level throughout this work.}
  \begin{tabular}{llccc}
  \hline
GRB & $z$ & $\text{T}_{90}$ (s) & $\Gamma_{\gamma}$ & $D_{\text{L}}$ (Mpc) \\
\hline
051221A & 0.55 & $1.4 \pm 0.2$ & $1.39 \pm 0.06$ & 3275.3 \\
050724 & 0.2576 & $3.0 \pm 1.0$ & $1.89 \pm 0.22$ & 1349.1 \\
060313 & (0.72) & $0.7 \pm 0.1$ & $0.71 \pm 0.07$ & 4536.9 \\
060801 & 1.13 & $0.5 \pm 0.1$ & $0.47 \pm 0.24$ & 7881.9 \\
070724A & 0.46 & $0.4 \pm 0.04$ & $1.81 \pm 0.33$ & 2645.9 \\ 
070809 & 0.219 & $1.3 \pm 0.1$ & $1.69 \pm 0.22$ & 1123.2 \\
080426 & (0.72) & $1.7 \pm 0.4$ & $1.98 \pm 0.13$ & 4536.9 \\
080905A & 0.122 & $1.0 \pm 0.1$ & $0.85 \pm 0.24$ & 590.7 \\
080919 & (0.72) & $0.6 \pm 0.1$ & $1.10 \pm 0.26$ & 4536.9 \\
090426 & 2.6 & $1.2 \pm 0.3$ & $1.93 \pm 0.22$ & 21805.6 \\
090510 & 0.9 & $0.3 \pm 0.1$ & $0.98 \pm 0.20$ & 5959.6 \\
090515 & (0.72) & $0.04 \pm 0.02$ & $1.60 \pm 0.20$ & 4536.9 \\
100117A & 0.92 & $0.30 \pm 0.05$ & $0.88 \pm 0.22$ & 6122.5 \\
100702A & (0.72) & $0.16 \pm 0.03$ & $1.54 \pm 0.15$ & 4521.6 \\
101219A & 0.718 & $0.6 \pm 0.2$ & $0.63 \pm 0.09$ & 4536.9 \\
111020A & (0.72) & $0.4 \pm 0.09$ & $1.37 \pm 0.26$ & 4536.9 \\
120305A & (0.72) & $0.10 \pm 0.02$ & $1.00 \pm 0.09$ & 4536.9 \\
120521A & (0.72) & $0.45 \pm 0.08$ & $0.98 \pm 0.22$ & 4536.9 \\
051210 & (0.72) & $1.4 \pm 0.2$ & $1.10 \pm 0.30$ & 4536.9 \\
061201 & 0.111 & $0.8 \pm 0.1$ & $0.81 \pm 0.15$ & 533.7 \\
070714A & (0.72) & $2.0 \pm 0.3$ & $2.60 \pm 0.20$ & 4536.9 \\
110112A & (0.72) & $0.5 \pm 0.1$ & $2.14 \pm 0.46$ & 4536.9 \\
111117A & (0.72) & $0.47 \pm 0.09$ & $0.65 \pm 0.22$ & 4536.9 \\
111121A & (0.72) & $\sim 0.47$ & $1.66 \pm 0.12$ & 4536.9 \\
100625A & 0.425 & $0.33 \pm 0.03$ & $0.90 \pm 0.10$ & 2409.3 \\
\hline
\end{tabular}
\label{tab:grbpars}
\end{table*}

An important distinction between our fitting procedure and that of \cite{rowl13} [see also \cite{lu15}] is that we do not include an explicit power-law component in the overall luminosity profile. Such a component is expected on physical grounds, namely that there is some relic contribution from the prompt emission \citep{obr06} and that the magnetar input, while dominant, is but an extra contribution to the overall luminosity, which may include fireball \citep{sarin19} and external shock \citep{dall11} pieces. We opt not to fit for the tail of the prompt emission, since this is largely independent from the properties of the host magnetar and it is therefore unclear if it is physically meaningful to fit for different prompt tails in our models. This choice leads to a fairer and more direct comparison between magnetar models.

{For each fit detailed herein, we use a Monte Carlo method with $\lesssim 10^{8}$ points sampled from the parameter space, which is 5-dimensional (spanned by $10^{-1} \leq B_{p,15} \leq 2 \times 10^{2} , 10^{-1} \leq P_{0,-3} \leq 10^{2}, 10^{-6} \leq \epsilon_{p} \leq 10^{-1},-1 \leq \alpha_{0} \leq 1$, and $-1 \leq k \leq 1$) and 2-dimensional (spanned by $10^{-1} \leq B_{p,15} \leq 2 \times 10^{2}$ and $10^{-1} \leq P_{0,-3} \leq 10^{2}$) for precessing and orthogonal fits, respectively. The outer boundaries are chosen to lie within the regime of physically acceptable magnetar parameters: in some cases, larger polar field strengths lead to a smaller mean-square error, though we reject these minima because they lead to \emph{internal} field strengths that exceed the Virial limit (see Sec. 4.4).}

\subsection{Akaike information criterion}

It is important to note that the parameter space for the precessing model is higher dimensional than the parameter space for the orthogonal model. In particular, since the orthogonal rotator is but a special case of the precessing model (with $\epsilon_{p} = 0, \alpha_{0} = 1$, and $k=0$), we should always obtain a better fit (in the Levenberg-Macquardt sense) in the former case, irrespective of whether the additional parameters are physically relevant or not. 

To quantify the extent to which precession-related improvements are not factitious artifacts but rather represent meaningful physics, we can calculate the associated Akaike information criterion (AIC) of each fit. From the Monte Carlo output, the AIC is computed via \citep{ak74}
\begin{equation} \label{eq:aicn}
\text{AIC} = 2 \left( N - \underset{m}{\text{max }} \hat{\ell} \right),
\end{equation}
for $N$ parameters to be estimated, where $\hat{\ell}$ is the log-likelihood function of the model. Given a set of candidate models for some data, the one with the minimum AIC value is preferred \citep{akaike}. In particular, a lower AIC implies that the information content of the fit is higher and, importantly, that the fit is preferred even if the mean-square residuals are worse.

As previously noted, the precessing model contains 3 additional parameters ($\alpha_{0}, k, \epsilon_{p}$) over the orthogonal model. In the event that the data strongly favours an orthogonal fit (see Appendix A), so that all precessing parameters effectively vanish within $L_{p}$, we see, from expression \eqref{eq:aicn}, that the AIC numbers predicted by the precessing models will be larger than those for the orthogonal model by six, i.e. $\text{AIC}_{p} - \text{AIC}_{\perp} = 6$. As such, in cases where the fits predict AIC numbers that are roughly within six of each other, the orthogonal model should be considered as heavily favoured. The extent of preference scales with the difference in the AIC numbers. Some tests on the robustness of the AIC are presented in Appendix A.

\section{Magnetar parameter fits}

In this work, we fit directly to the observed X-ray flux $F$ in the 0.3$-$10 keV band, which is related to the afterglow luminosity $L$ through 
\begin{equation} \label{eq:fluxlum}
L(t) = 4 \pi D_{L}^2 F(t) K(z),
\end{equation}
for luminosity distance $D_{L}$ (see Tab. \ref{tab:grbpars}) and cosmological $K$-correction factor $K(z) \approx (1+z)^{\Gamma_{\gamma}-2}$ \citep{bloom01}. Note that expression \eqref{eq:fluxlum} assumes uncollimated emission; for collimated emission, the right-hand side of \eqref{eq:fluxlum} is multiplied by a factor $1 - \cos \theta_{\text{jet}}$, where $\theta_{\text{jet}}$ is the half-opening angle of the jet \citep{pesc15,stratta18}. The parameters inferred for a collimated outflow are generally more extreme, since small opening angles, as predicted by \cite{mur17} from numerical simulations of jet break out, may lead to a factor $\sim 10$ {decrease} in the predicted $L$ [hence a factor $\sim 10^{0.5}$ increase in $B_{p}$ {for fixed $\tau_{\text{sd}}$; see expression (1) of \cite{rowl17}}]. The parameter $\theta_{\text{jet}}$ is, however, degenerate with the efficiency $\eta$, which is why we ignore it here.

\begin{figure*}
\includegraphics[width=\textwidth]{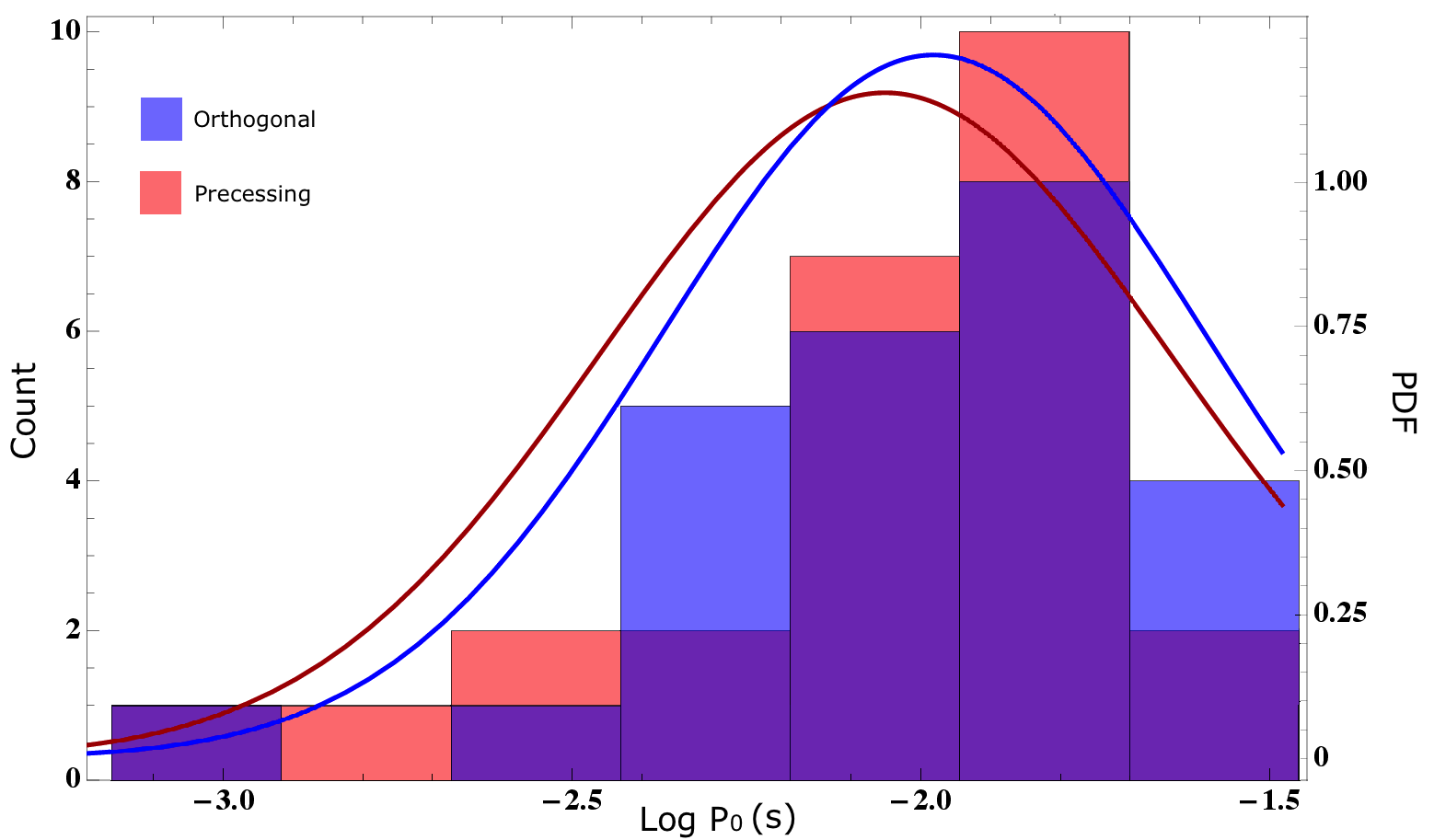} 
\caption{Histogram (count on left y-axis) with overplotted normal fitting (PDF on right y-axis) for the spin periods at birth, $P_{0}$, for precessing (red) and orthogonal (blue) fits.} \label{fig:pdist}
\end{figure*}
\begin{figure*}
\includegraphics[width=\textwidth]{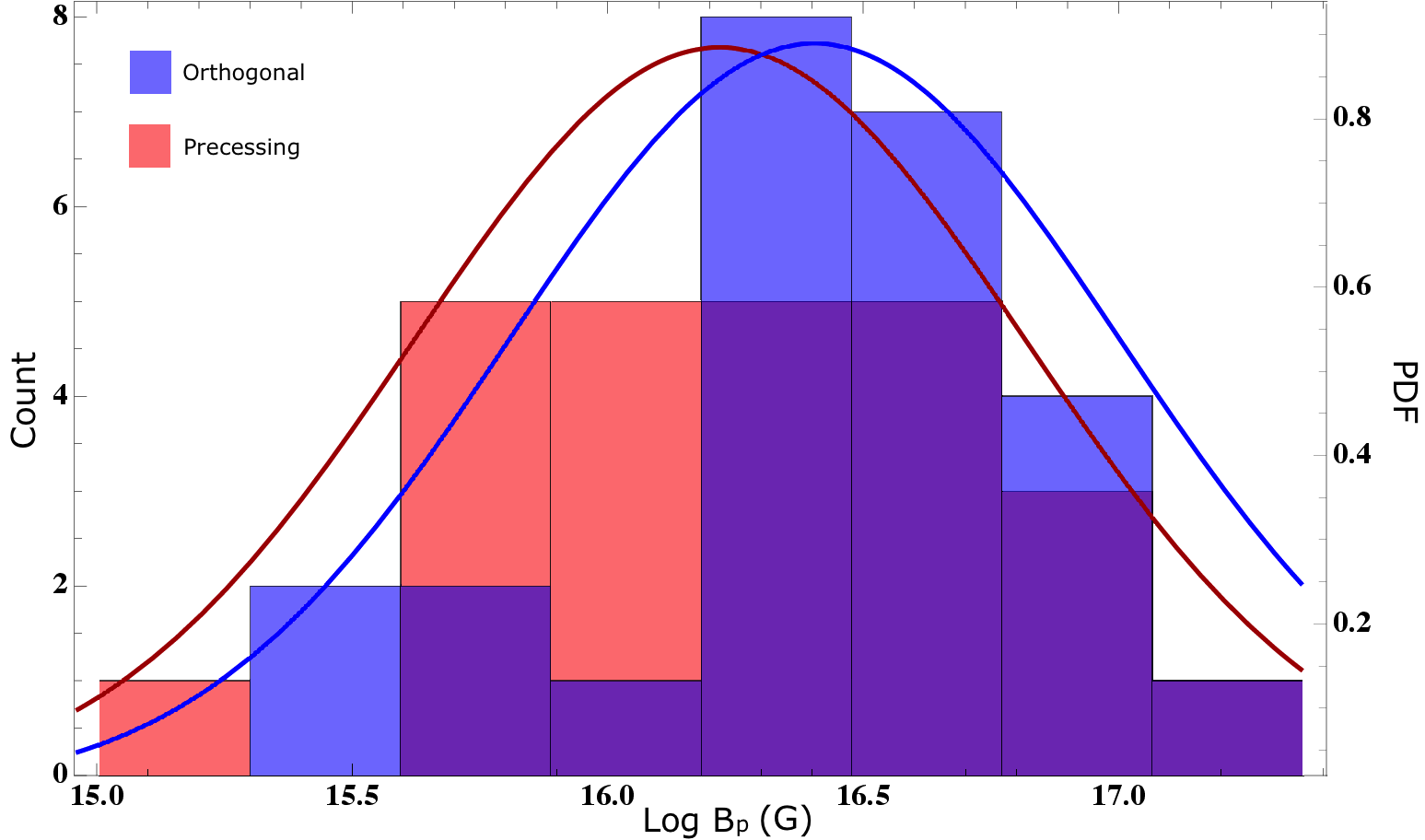} 
\caption{Histogram (count on left y-axis) with overplotted normal fitting (PDF on right y-axis) for the polar field strengths, $B_{p}$, for precessing (red) and orthogonal (blue) fits.} \label{fig:bdist}
\end{figure*}

Tables \ref{tab:precfits} and \ref{tab:orthfits} report the magnetar parameters inferred from precessing, oblique rotator and orthogonal rotator fits, respectively. We emphasise again that the fits obtained herein assume perfect efficiency $(\eta = 1)$, and we further take fixed values of the mass [$M = 2.0 M_{\odot}$; consistent with observations of PSR J0740$+$6620 \citep{crom19}] and radius ($R = 10^{6} \text{ cm}$), appropriate for a highly-compact NS born out of a merger. In general, the spin periods at birth scale as $P_{0} \propto M_{2.0}^{1/2} R_{6} \eta^{1/2}$, the polar field strengths as $B_{p} \propto M_{2.0} R_{6}^{-1} \eta^{1/2}$, and the ellipticities as $\epsilon_{p} \propto M_{2.0}^{1/2} R_{6}  \eta^{1/2}$. The distributions of spin periods at birth and polar field strengths are shown in Figures \ref{fig:pdist} and \ref{fig:bdist}, respectively.

\begin{table*}
\caption{Magnetar parameters inferred from precessing, oblique rotator fits to Swift-XRT data pertaining to the GRBs listed in Tab. \ref{tab:grbpars}. As discussed in Sec. 3, spin periods scale as $P_{0} \propto M_{2.0}^{1/2} R_{6} \eta^{1/2}$, polar field strengths as $B_{p} \propto M_{2.0} R_{6}^{-1} \eta^{1/2}$, and ellipticities as $\epsilon_{p} \propto M_{2.0}^{1/2} R_{6} \eta^{1/2}$. AIC numbers are computed from equation \eqref{eq:aicn}. Collapse times are quoted from Rowlinson et al. (2013).}
  \begin{tabular}{llccccccc}
  \hline
GRB & $P_{0}$ (ms) & $B_{p}$ ($10^{15}$ G) & $\ep$ ($\times 10^{-3}$) & $\Lambda$ ($\times 10^{-1}$) & Collapse time (s) & $\text{AIC}_{p}$  \\
\hline
051221A & $16.3 \pm 0.21$ & $5.67 \pm 0.1$ & $0.43 \pm 0.01$ & $0.70 \pm 0.02$ & - & -31.8 \\
050724 & $7.33 \pm 0.19$ & $60.6 \pm 1.6$ & $0.10 \pm 0.01$ & $3.85^{+0.02}_{-0.29}$ & - & 321 \\
060313 & $11.6 \pm 0.34$ & $14.7 \pm 0.7$ & $0.011 \pm 0.001$ & $3.83^{+0.04}_{-0.5}$ & - & -250 \\
060801 & $13.8 \pm 1.3$ & $70.7 \pm 6.3$  & $16.0 \pm 1.5$ & $1.86 \pm 0.44$ & 326 & -2.49  \\
070724A & $9.4 \pm 0.6$ & $22.2 \pm 1.5$ & $0.12 \pm 0.1$ & $3.60^{+0.10}_{-0.65}$ & 90 & 537  \\
070809 & $6.65 \pm 0.15$  & $4.0 \pm 0.1$ & $0.026 \pm 0.01$ & $2.51 \pm 0.14$ & - & -133  \\
080426 & $14.7 \pm 0.5$ & $13.1 \pm 0.5$ & $7.03 \pm 0.25$ & $0.26 \pm 0.02$ & - & -136  \\
080905A & $19.4 \pm 0.3$ & $137 \pm 2$ & $36.7 \pm 0.5$ & $2.33 \pm 0.08$ & 274 & 23.1 \\
080919 & $7.8 \pm 3.4$ & $20.0 \pm 9.0$ & $2.36 \pm 1.14$ & $1.29 \pm 1.27$ & 421 & 15.6 \\
090426 & $3.50 \pm 0.6$ & $4.8 \pm 1.0$ & $(1.32 \pm 0.29)$ $\times 10^{-3}$  & $3.81^{+0.07}_{-2.2}$ & - & -156 \\
090510 & $8.10 \pm 0.5$ & $14.5 \pm 0.9$ & $9.02 \pm 0.58$ & $0.25 \pm 0.04$ & - & 69.8 \\
090515 & $3.89 \pm 0.51$ & $5.78 \pm 1.3$ & $(6.55 \pm 0.85)$ $\times 10^{-3}$ & $3.65^{+0.23}_{-2.1}$ & 175 & 170 \\
100117A & $6.14 \pm 0.36$ & $15.0 \pm 1.1$ & $5.86 \pm 0.44$ & $0.40 \pm 0.06$ & - & 891 \\
100702A & $1.75^{+3.25}_{-1.25}$ & $57.2^{+172}_{-51}$ & $0.11^{+0.27}_{-0.09}$ & $3.85^{+0.03}_{-3.7}$ & 178 & 560   \\
101219A & $0.90 \pm 0.02$ & $1.62 \pm 0.04$ & $2.04 \pm 0.05$ & $0.015 \pm 0.001$ & 138 & 28.8   \\
111020A & $15.2 \pm 1.2$ & $12.0 \pm 1.0$ & $0.112 \pm 0.1$ & $3.1 \pm 0.6$ & - & -104   \\
120305A & $8.72 \pm 0.21$ & $40.9 \pm 1.0$ & $1.44 \pm 0.04$ & $3.01 \pm 0.19$ & 182 & 664   \\
120521A & $19.5^{+23.5}_{-16.2}$ & $34.2^{+40.8}_{-32.6}$ & $21.6^{+26.4}_{-20.1}$ & $0.53^{+1.37}_{-0.44}$ & 207 & 3.43   \\
051210 & $11.4 \pm 0.9$ & $41.4 \pm 3.4$ & $7.21 \pm 0.06$ & $1.60 \pm 0.29$ & 225 & 60.2   \\
061201 & $22.6 \pm 0.3$ & $42.0 \pm 1$ & $1.50 \pm 0.04$ & $3.01 \pm 0.19$ & - & -14.4  \\
070714A & $12.2 \pm 1.5$ & $6.3 \pm 0.8$ & $14.4 \pm 1.7$ & $0.03 \pm 0.01$ & - & -77.9   \\
110112A & $21.3 \pm 2.8$ & $24.8 \pm 3.2$ & $0.16 \pm 0.02$ & $3.6^{+0.22}_{-1.3}$ & - & -75.4   \\
111117A & $17.3 \pm 3.7$ & $62.1 \pm 15.9$ & $10.1 \pm 2.5$ & $2.05 \pm 1.25$ & - & 1.31  \\
111121A & $2.67 \pm 0.09$ & $24.8 \pm 0.8$ & $0.13 \pm 0.01$ & $3.62^{+0.15}_{-0.32}$ & - & 861   \\
100625A & $10.3 \pm 0.5$ & $21.7 \pm 1.2$ & $9.8 \pm 0.5$ & $0.48 \pm 0.06$ & - & 4.97   \\
\hline
\end{tabular}
\label{tab:precfits}
\end{table*}

\begin{table}
\caption{Magnetar parameters inferred from orthogonal rotator fits to Swift-XRT data pertaining to the GRBs listed in Tab. \ref{tab:grbpars}. As discussed in Sec. 3, spin periods scale as $P_{0} \propto M_{2.0}^{1/2} R_{6} \eta^{1/2}$ and polar field strengths as $B_{p} \propto M_{2.0} R_{6}^{-1} \eta^{1/2}$. AIC numbers are computed from equation \eqref{eq:aicn}. Those AIC numbers which are either marginally larger or smaller than the corresponding precessing fits, thereby indicating that there is weak or no evidence for a precessing source, respectively, are shown with superscript daggers.}
\centering
  \begin{tabular}{llccccc}
  \hline
GRB & $P_{0}$ (ms) & $B_{p}$ ($10^{15}$ G) & $\text{AIC}_{\perp}$ \\
\hline
051221A & $16.7 \pm 0.4$ & $5.26 \pm 0.2$ & -9.53  \\
050724 & $6.65 \pm 0.17$ & $69.9 \pm 1.8$ & 415   \\
060313 & $12.7 \pm 0.32$ & $14.1 \pm 0.5$ & -196  \\
060801 & $13.6 \pm 1.8$ & $49.9 \pm 10.1$ & 15.5   \\
070724A & $10.9 \pm 1.8$ & $52.7 \pm 15.3$ & 2124    \\
070809 & $6.45 \pm 0.24$ & $2.97 \pm 0.19$ & -124$^{\dag}$   \\
080426 & $15.6 \pm 0.7$ & $20.6 \pm 1.5$ & -125$^{\dag}$  \\
080905A & $17.2 \pm 3.8$ & $186 \pm 58$ & 47.5   \\
080919 & $8.32 \pm 3.0$ & $45.9 \pm 25.2$ & 9.92$^{\dag}$   \\
090426 & $3.91 \pm 0.69$ & $6.2 \pm 1.6$ & -128  \\
090510 & $7.83 \pm 0.47$ & $17.5 \pm 1.2$ & 123   \\
090515 & $5.51 \pm 0.3$ & $20.8 \pm 1.1$ & 341   \\
100117A & $6.32 \pm 0.48$ & $22.5 \pm 1.7$ & 1461   \\
100702A & $5.84 \pm 0.25$ & $27.3 \pm 1.4$ & 863  \\
101219A & $0.92 \pm 0.05$ & $2.16 \pm 0.24$ & 81.9    \\
111020A & $26.0 \pm 2.5$ & $23.5 \pm 2.8$ & -59.1  \\
120305A & $8.29 \pm 0.22$ & $40.0 \pm 1.2$ & 1917   \\
120521A & $23.5 \pm 2.5$ & $76.2 \pm 7.8$ & 0.60$^{\dag}$  \\
051210 & $10.8 \pm 0.9$ & $66.1 \pm 5.9$ & 203  \\
061201 & $23.5 \pm 1.4$ & $53 \pm 6$ & -9.80$^{\dag}$ \\
070714A & $18.7 \pm 1.8$ & $17.2 \pm 2.5$ & -69.9$^{\dag}$  \\
110112A & $26.0 \pm 4.0$ & $33.0 \pm 5.0$ & -61.7$^{\dag}$   \\
111117A & $17.7 \pm 15.3$ & $76.8^{+85.2}_{-75.1}$ & -4.30$^{\dag}$  \\
111121A & $3.15 \pm 0.1$ & $28.6 \pm 0.9$ & 1271  \\
100625A & $11.5 \pm 6.5$ & $44.3 \pm 35.7$ & -0.82$^{\dag}$   \\
\hline
\end{tabular}
\label{tab:orthfits}
\end{table}

\subsection{Number of systems favouring precession}

Based on the AIC (Sec. 3.1), we find that 16 out of the 25 $(64\%$) SGRBs within our sample favour a precessing fit. This constitutes the main result of this work: including the effects of precession not only reduces the mean-square residuals of the fits (by over an order of magnitude in some cases, e.g., GRB 100625A), but also leads to an information-theoretically preferred model in $\gtrsim 60\%$ of cases. This result, while the first of its kind to the authors' knowledge, is not entirely surprising. It is generally expected that newborn NSs undergo precession, as violent activity at birth resets any pre-existing relationship between the Euler angles of the system \citep{thom93,melatos00,lander18}. 

Interestingly, the objects that display the strongest evidence for precession (in the sense that the difference between the AIC numbers is greatest) are the supramassive systems GRB 070724A and GRB 120305A; the penultimate column in Tab. \ref{tab:precfits} lists the collapse time of the associated remnant based on the analysis of \cite{rowl13}. In particular, those sources which display a sharp drop (temporal index $\ll -2$) in the afterglow luminosity indicate a cessation of the energy injection, which can be associated with collapse to a black hole \citep{rowl10,lasky14}. The connection between supramassive stars and strong precession is also expected on physical grounds, since a higher rotational velocity, as necessary to stave off collapse, correlates to a higher precession frequency [equation \eqref{eq:precfreq}].

\subsection{Spin frequencies and magnetic fields}

Denoting mean values by an overhead bar, we find $\bar{P}^{p}_{0,-3}/\left(M_{2.0}^{1/2} R_{6} \eta^{1/2}\right) = 8.61$ (see Figure \ref{fig:pdist}) and $\bar{B}^{p}_{p,15}/\left(M_{2.0} R_{6}^{-1} \eta^{1/2}\right) = 19.0$ (see Figure \ref{fig:bdist}) for the precessing fits. While the magnetic field strength is on the high-end, numerical simulations of dynamo activity \citep{dunc1,miralles02}, magnetorotational amplification \citep{sawai13}, and the Kelvin-Helmholtz instability \citep{ross03,kiu15} suggest that field strengths of this order may be easily achieved in a post-merger remnant. For the subset of SGRBs that exhibit precursor flashes [e.g., GRB 090510\footnote{Note that, because we assume the magnetospheric $\delta$ is fixed at unity, the values inferred here are slightly different ($\lesssim 15\%$ difference in $B_{p}$ and $P_{0}$) than those reported in \cite{suv20}.} \citep{coppin20}], \cite{suvk20} suggested that the pre-merger stars may themselves be highly-magnetised so as to explain the lack of thermal emission in the precursors \citep{zhong19}. Through flux freezing, this may also help to explain the ultra-magnetisation of the remnant.

Although sensitive to the magnetization parameter of the stellar wind, \cite{xiao19} found that $\eta \approx 5 \times 10^{-2}$ near the beginning of the extended emission phase for GRBs 100615A and 150910A. While these two latter GRBs are not in our sample, taking this value of $\eta$ as canonical we find more modest values for the mean magnetic field strength, $\bar{B}^{p}_{p} = 4.26 \times 10^{15} \text{ G}$, and spin frequency at birth, $\bar{\nu}_{p,0} = 520 \text{ Hz}$. However, this value of $\eta$ is likely inappropriate for all magnetars within our sample, as it depends on the bulk Lorentz factor of the wind and is sensitive to, amongst other things, the extent of baryon pollution in the environment surrounding the remnant \citep{thom04,lee07,mur14}.

In particular, the most rapidly rotating magnetar we find within our sample corresponds to (supramassive) GRB 101219A, which, for a value $\eta \gtrsim 0.5$, gives $\nu^{\text{101219A}}_{p,0} \lesssim 1571 \text{ Hz}$. This value is comparable to the Keplerian break-up (mass shedding) limit $\nu_{\text{K}}$; from numerical simulations, \cite{haen09} found an EOS-independent empirical formula for the break-up limit in the form $\nu_{\text{K}} = \sqrt{2} C M_{2.0}^{1/2} R_{6}^{-3/2}$, where $C = 1.08$ kHz for a hadronic star and $C = 1.15$ kHz for a strange star. The closeness between the break-up limit and the fitted spin frequency is consistent with the object being supramassive, as the collapse time $t_{\text{col}} \sim 178$ s \citep{rowl13}. Furthermore, using a variety of equations of state, \cite{li16} found that the narrow break times observed in some SGRBs with internal plateaus are suggestive of supra-massive quark stars rather than NSs. For these objects, $C \approx 1.15$ kHz is appropriate, and the mass-shedding frequency matches well with the inferred spin value for $\eta \approx 0.5$, though it would be difficult to accommodate a meta-stable magnetar with $\eta \lesssim 0.5$. In any case, rapidly rotating and massive stars are likely subject to the Chandrasekhar-Friedman-Schutz (CFS) instability for $f$- and/or $r$-modes \citep{kru20,kru20b}, which would lead to episodes of accelerated spindown and enhanced gravitational-wave emission that may be discernible in the astrophysical data \citep{don15}.

We remark that while NSs with $\gtrsim$ kHz spin frequencies have not been observed directly, \cite{yamas20} recently suggested that successive pulses in FRB 181112 hint at a magnetar source with $P_{0} \sim 0.8$ ms, a value that is not unusual for remnants in NSNS merger simulations \citep{giac13,ciolfi17,radice18}.

For the orthogonal fits, we find $\bar{B}^{\perp}_{p,15}/\left(M_{2.0} R_{6}^{-1} \eta^{1/2}\right) = 26.2$ and $\bar{P}^{\perp}_{0,-3}/\left(M_{2.0}^{1/2} R_{6} \eta^{1/2}\right) = 9.85$, which are both larger than the precessing fits, implying that the orthogonal fits favour slower, but more magnetised, stars. In general, our predictions for the orthogonal models differ somewhat from those of \cite{rowl13} and \cite{lu15}. The reasons for this are fourfold, namely because (i) we use slightly different mass $(M = 2.0 M_{\odot})$ and moment of inertia $(I_{0} = 2 M R^2/5)$ normalisations, (ii) our selection criteria for `best-fit' differs, as we assume the best model is the one that minimises the least-squares error, (iii) we do not include a power-law component to account for the tail of the prompt emission (which leads to a slight overestimation for the polar field strength in general), and (iv) we assume a different cosmology, using updated values from \cite{planck18}. In general, however, our fits are mostly consistent with those of previous authors in the case of orthogonal rotators.

\subsection{Ellipticities}

\begin{figure}
\includegraphics[width=0.473\textwidth]{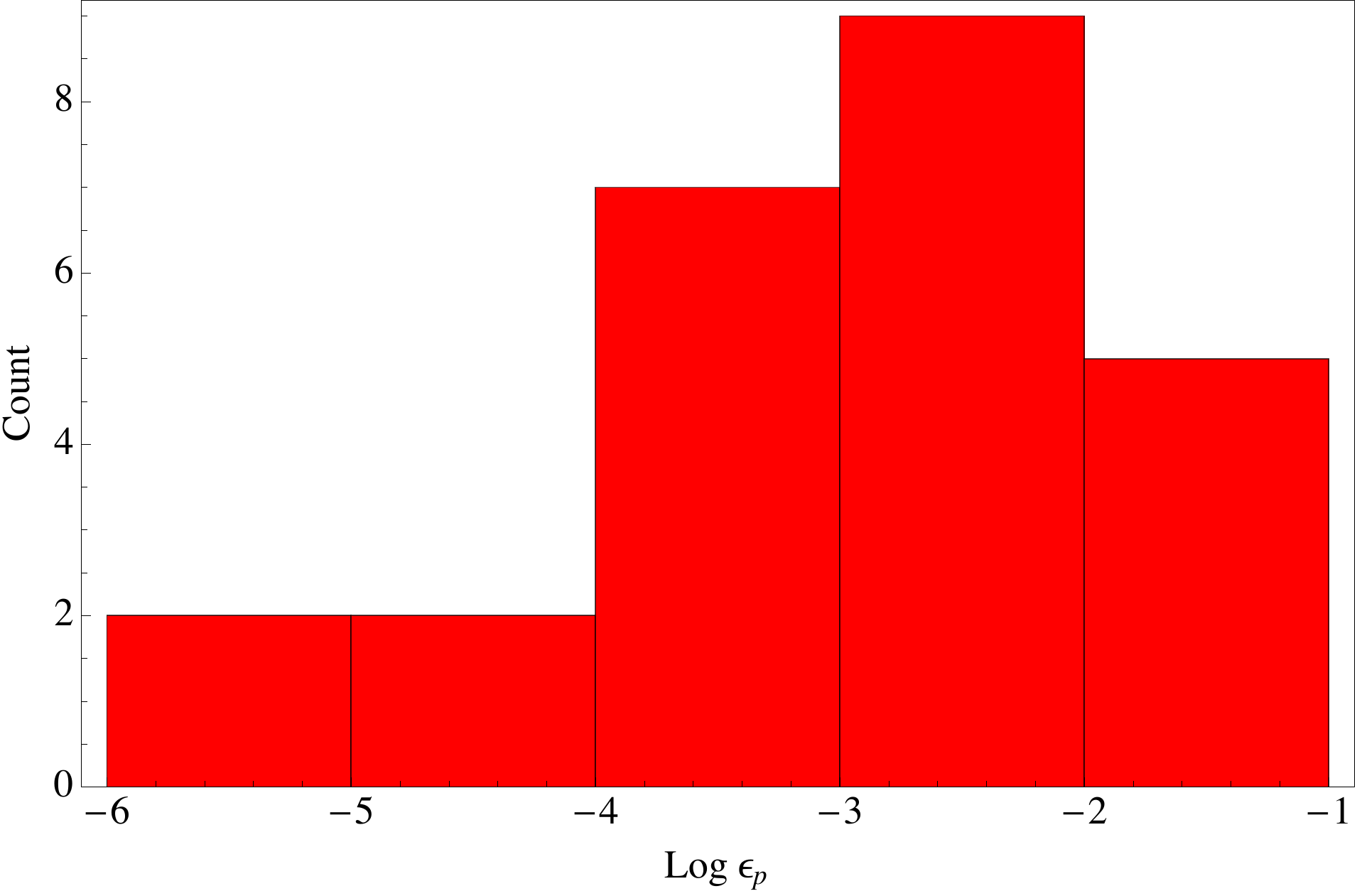} 
\caption{Histogram for the spin-misaligned ellipticities of the precessing fits shown in Table \ref{tab:precfits}.} \label{fig:ellips}
\end{figure}

One aspect of the precessing fits, not available for the orthogonal fits (though see below), is an independent inference of the stellar ellipticity via equation \eqref{eq:precfreq}. We find $\bar{\epsilon}_{p}/\left(M_{2.0}^{1/2} R_{6} \eta^{1/2}\right) = 7.73 \times 10^{-4}$, which is consistent with theoretical predictions for internal deformations of newborn magnetars \citep{hask08,mast11,dall14} even though we assume no \emph{a priori} relationship between $\epsilon_{p}$ and the other stellar parameters. From the values of $\epsilon_{p}$ and $P_{0}$, we find that the precession periods of the magnetars in our sample span the range $\sim 0.4 - 3 \times 10^{3}$ s. Figure \ref{fig:ellips} illustrates the distribution of stellar ellipticities collated in Table \ref{tab:precfits}.

We find that those objects with the largest ellipticities predicted by our fits tend to also be those that collapse (e.g., $\epsilon_{p} = 3.7 \times 10^{-2}$ for GRB 080905A, which collapses $\sim 274$ seconds after formation). {Two notable exceptions are GRB 070714A and GRB 100625A, which house large ellipticities ($\gtrsim 10^{-2}$) but do not obviously collapse [see however \cite{lu15}], though these latter fits are disfavoured by the AIC \citep{akaike} since $\text{AIC}_{p} = -77.9$ is only marginally smaller than $\text{AIC}_{\perp} = -69.9$ for GRB 070714A and $\text{AIC}_{\perp} < \text{AIC}_{p}$ for GRB 100625A}. On the other hand, GRB 090515 shows clear evidence for collapse and yet we predict a relatively mild ellipticity, $\epsilon_{p} \sim 7 \times 10^{-6}$, perhaps indicating that differential rotation was quenched early on in this object's life, so as to suppress the growth of the toroidal field (see Sec. 4.4) \citep{shap00}.

In general, however, the fact that the most deformed stars are also those that collapse is not entirely surprising. In the supramassive scenario, the star is supported by centrifugal forces which halt the collapse that would otherwise take place if the star were not rapidly rotating. For uniformly\footnote{This limit can be pushed considerably further if the star is hypermassive and supported by differential rotation; \cite{weih18} found that $M \gtrsim 1.54 M_{\text{max}}$ is required to instigate collapse. However, although dependent on the rate of angular momentum dissipation, collapse within $\lesssim$ seconds would be expected \citep{faber12}, which is too short to account for X-ray afterglows.} rotating remnants, \cite{breu18} [see also \cite{rezz18}] found that the mass limit for supramassive stars relative to non-rotating ones is $M \sim 1.2 M_{\text{max}}$. However, even if the star is somewhat below this maximum threshold, convection may drive localised regions within the core to become overdense, especially if there is an ultra-strong internal magnetic field\footnote{{Equipartition-level ($B_{\text{int}} \gtrsim 10^{18} \text{ G}$) magnetic field strengths can also help to push up the maximum mass threshold by $\lesssim 30\%$ \citep{boc95}.}}, which could lead to hydrodynamic instabilities that assist in the collapse. Large ellipticities $\epsilon_{p} \gtrsim 10^{-2}$, which imply the existence of highly-anisotropic density gradients, may therefore accelerate the collapse of an object that would have otherwise been stable for periods of time longer than the Swift-XRT window. This point will be investigated elsewhere.


\subsection{Toroidal fields}

A non-zero value for the spin-misaligned ellipticity $\epsilon_{p}$ implies an internal deformation within the star. As such, comparing the fitted value with a hypothetical value coming from some canonical model of magnetic deformations, we can infer some properties of the internal magnetic field.

For simplicity, suppose that the stellar magnetic field is predominantly dipolar [though cf. \cite{ciolfi09,suvk19}] and contains both poloidal and toroidal components. The relative strength between these two components is largely unknown. At least in certain systems, however, there is observational evidence for the existence of a strong toroidal component. In particular, pulse fractions in surface X-ray emissions \citep{shab12} and cyclotron resonant scattering features \citep{thom02} suggest that local magnetic field strengths may greatly exceed [$\gtrsim 3$ orders of magnitude in the case of 1E 1207.4-5209 \citep{pav02}] the global values inferred from spindown alone. Furthermore, while purely poloidal or toroidal fields are known to be unstable \citep{wri73,tay73}, \cite{brai09} has argued that predominantly toroidal fields may be stable under some circumstances in NSs [see also \cite{ciolfi13,mosta20}]. At least in the early stages of a magnetar's life, the winding of field lines due to dynamo activity is expected to generate a strong azimuthal component \citep{dunc1,miralles02}, which may be further amplified by magnetorotational \citep{sawai13} or Kelvin-Helmholtz \citep{ross03,kiu15} instabilities.

For a non-barotropic star with a parabolic (Tolman VII) density profile, \cite{mast11} [see also \cite{hask08,dall14}] found that the ellipticity induced by magnetic deformations reads
\begin{equation} \label{eq:epsilon}
 \epsilon_{B}  = 2.93 \times 10^{-6} B_{p,15}^2 M_{2.0}^{-2} R_{6}^{4} \left( 1 - \frac {0.389} {\Lambda} \right),
\end{equation}
where $\Lambda$ represents the poloidal-to-toroidal magnetic energy ratio (so that, e.g., a value of $\Lambda = 1$ indicates a purely poloidal magnetic field). Using expression \eqref{eq:epsilon}, we can estimate the value of $\Lambda$ for any particular GRB candidate by equating $\epsilon_{B} = \epsilon_{p}$ using the $B_{p}$ values obtained from the fitting procedure. Note that for $\Lambda < 0.389$, the deformation tends to drive the star towards a \emph{prolate} rather than \emph{oblate} shape \citep{mast11}, as expected of toroidal fields. However, the star is almost certainly still geometrically oblate since the rotational oblateness, while not instigating precession, tends to dominate (see Footnote 2). 

\begin{figure}
\includegraphics[width=0.473\textwidth]{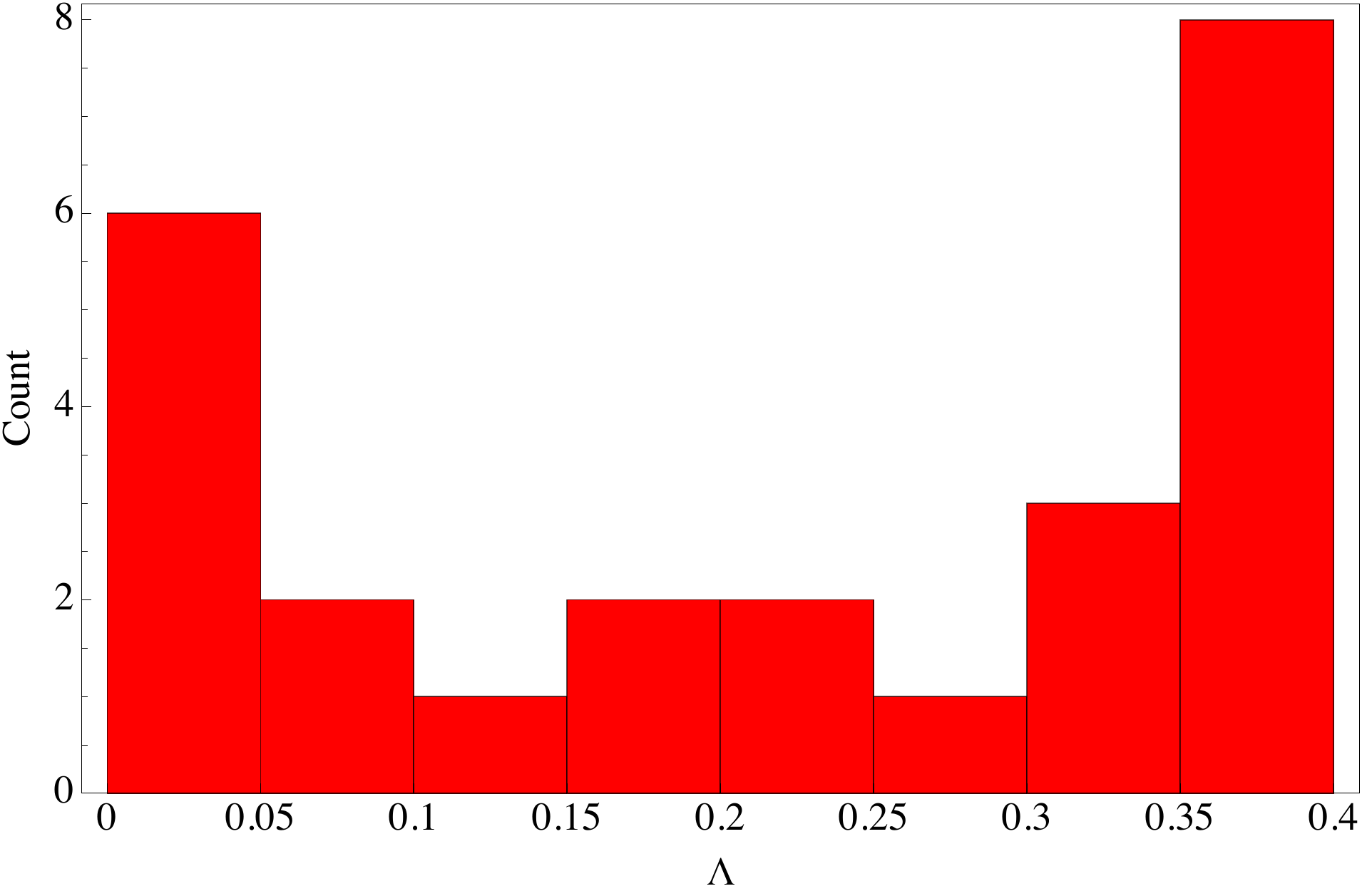} 
\caption{Histogram for the poloidal-to-toroidal energy ratio, $\Lambda$, estimated by equating the ellipticities shown in Fig. \ref{fig:ellips} with those from expression \eqref{eq:epsilon} derived by \protect\cite{mast11}. } \label{fig:lambdas}
\end{figure}

Figure \ref{fig:lambdas} shows the values of $\Lambda$ necessary to match $\epsilon_{B}$ from equation \eqref{eq:epsilon} with the aforementioned ellipticity values. The values of $\Lambda$ we find span the range $1.5 \times 10^{-3} \leq \Lambda \leq 3.85 \times 10^{-1}$, with lower limit coming from GRB 101219A and upper limit from GRBs 050724 and 100702A. {We caution the reader that for our modelling based on the analytic study of \cite{mast11}, the maximum \emph{internal} field strength predicted for our parameters (Tab. \ref{tab:precfits}) occurs within the equatorial torus of (supramassive) GRB 080905A, where we find $\Lambda = 0.23$ and $B_{\text{int,max}} / \sqrt{\eta} \sim 2.3 \times 10^{18}$ G. Such a value implies that the magnetic energy density matches the gravitational binding energy density (i.e., the Virial limit) in a small region bounded by the neutral line within the star for perfect efficiency $\eta=1$ \citep{latt07}, and therefore magnetic forces must be treated nonlinearly in this particular case \citep{boc95}. Assuming instead $\eta \lesssim 5 \times 10^{-2}$, as inferred by \cite{xiao19} for some systems, our prediction for the \emph{maximum} field strength drops to $B_{\text{int,max}} \lesssim 5 \times 10^{17}$ G, and it is reasonable to treat the Lorentz force as a perturbation relative to the hydrostatic force \citep{mast11}. For all other GRBs in our sample, the maximum field strengths we find are below the Virial limit even for $\eta = 1$. Additionally noting that $\Lambda \gtrsim 10^{-3}$ for some cases considered here, the stability analyses of \cite{akgun13} and \cite{herb17} imply that our magnetic fields may suffer from interchange and kink instabilities unless the stars are strongly stratified.}

While the distribution of stellar ellipticities {clusters around $\epsilon_{p} \sim 10^{-3}$}, the poloidal-to-toroidal strengths appear to follow a roughly bi-modal distribution. For instance, eight candidates (050724, 060313, 070724A, 090426, 090515, 100702A, 110112A, 111121A) out of our sample favour an almost energetically-equal partition with $\Lambda \approx 0.38$. In this case, the prolate deformation induced by the poloidal field is counterbalanced by the oblate deformation induced by the toroidal field, thereby leading to a relatively small ellipticity \citep{mast11}. On the other hand, eight candidates (051221A, 070714A, 080426, 090510, 100117A, 100625A, 101219A, 120521A) favour a predominantly toroidal configuration with $\Lambda \leq 0.1$ (i.e., at least $90\%$ of the magnetic energy is concentrated within the toroidal field). In the former case, one would expect low gravitational wave luminosities to be associated with the remnants, with the reverse situation applying in the latter case.

Indeed, a time-dependent mass quadrupole moment leads to gravitational radiation [see e.g. \cite{lasky16}], which we do not model here. In reality, both orthogonal and precessing rotators with internal deformations emit gravitational waves with an amplitude that is directly proportional to $\epsilon_{p}$. Although beyond the scope of this work, it would be interesting to include this spindown contribution and study the braking indices $n$ for those models predicted to house a strong toroidal component $(\Lambda \lesssim 0.1)$. As found by \cite{don15}, the absence of a comparable poloidal field suggests that spindown is likely dominated by gravitational-wave emission due to the bar-mode (CFS) instability, which is expected in rapidly rotating and massive NSs \citep{kru20,kru20b}. In general, a value of $n$ close to 5 indicates that a substantial gravitational-wave luminosity could be expected \citep{sarin18}. Using the relativistic shock model of \cite{dall11} which includes radiative losses [see also \cite{lasky17}], \cite{sarin20} found $n \gtrsim 4$ for GRB 111020A. This is consistent with our finding $\epsilon_{p} = 1.1 \times 10^{-4}$ and $B^{p}_{p,15} = 12.0$ for this object provided $\eta \ll 1$; for these values, neither electromagnetic nor gravitational braking would be expected to overly dominate the spindown profile at late times [see, e.g., equation (23) in \cite{mast11}].

By contrast, for example, \cite{sarin20} found that $n \approx 2$ for GRB 070809. For this braking index, gravitational-wave induced spindown is expected to be marginal (consistent with our finding $\epsilon_{p} \sim 2 \times 10^{-5}$ for this candidate). Moreover, a value of $n<3$ in general suggests that the magnetosphere may be twisted by the bubbling up of toroidal fluxes from the star \citep{par12,lasky17}, also consistent with a dynamical inclination angle for $\dot{\alpha} > 0$ and, by extension, a precessing model; in general, for $\lambda(\alpha) = 1 + \sin^2\alpha$, one has \citep{spit06,lai15}
\begin{equation}
n = 3 -  \frac {2 P \dot{\alpha} \sin \alpha \cos \alpha} {\dot{P} \left(1 + \sin^2 \alpha \right)}.
\end{equation}

Systems with $n<3$ may also indicate that the source suffered significant fallback accretion: super-Eddington accretion rates, $\dot{M}_{a} \lesssim 0.1 M_{\odot} \text{ s}^{-1}$, which can be expected shortly after the merger in some circumstances, can drastically modify the spindown evolution \citep{metz18}, {especially because the dipole moment may be reduced by a factor $\gtrsim 10$ within the first $\sim 10^{3}$ seconds of accretion via magnetic burial \citep{mel14,li20}}. The tilting of $n$ towards values $<3$ may, however, be combatted somewhat by the inducement of multipolarity due to field line compression \citep{suvm20} since $n \sim 2\ell+1$ for a pure $\ell$-pole \citep{pet19}. Incidentally, \cite{jin20} recently identified a blue kilonova counterpart to GRB 070809, providing further evidence for a magnetar central-engine model in this particular source. In any case, the ellipticities we find (Fig. \ref{fig:ellips}) are roughly consistent with those found by \cite{lasky16} for magnetar candidates in SGRBs (see Figure 2 therein), though we approach the problem from a different viewpoint.

\section{Discussion}

While there are some theoretical hurdles associated with newborn NSs producing the collimated outflows necessary for SGRBs \citep{ciolfi17,ciolfi18,cio20}, the birth of a stable or meta-stable magnetar may help to explain a number of features observed in some sources \citep{piro17}. In particular, the emergence of X-ray plateaus \citep{zhang01,fan06,tang19}, surviving well beyond the expected fallback timescale for black holes $(\lesssim$ seconds), and late-time X-ray flares \citep{camp06,mar11} strongly hint at a magnetar central-engine. Such a star will strongly precess in the early stages of its life if the Euler angles of the system are essentially stochastic relative to the pre-merger system, which may lead to a modulation of the X-ray luminosity as the source realigns over time, as has been observed directly in some systems \citep{hard91,usov92,marg08}. Based on an analysis of oscillations in the X-ray plateau, \cite{suv20} recently reported evidence for precession in GRB 090510. 

In this work, to analyse the statistical robustness of that result and to see whether precession may be an active component in other newborn magnetars, we fit an approximate expression for the spindown appropriate for a precessing, oblique rotator [equation \eqref{eq:genspinlum}] to X-ray lightcurve data of 25 SGRBs (see Tab. \ref{tab:grbpars}) that exhibit extended emission. Our sample consists of a subset (see Sec. 3) of those SGBRs where a possible magnetar progenitor was identified by \cite{rowl13} and \cite{lu15}. The results of the Monte Carlo fitting are summarised in Tab. \ref{tab:precfits}, which can be compared with the fits obtained using a luminosity profile appropriate for an orthogonal rotator [equation \eqref{eq:spiinlum}; Tab. \ref{tab:orthfits}]. 

We find that there is statistical evidence (determined via the AIC; see Sec. 3.1 and Appendix A) for precessional oscillations in 16 of the 25 $(\sim 60\%)$ SGRBs within our sample. This suggests that precession is a relatively generic feature of newborn NSs; a result which is expected on theoretical grounds \citep{thom93,melatos00,lander18} [see also \cite{link01,heyl02,post13}]. The existence of precession in magnetars specifically is supported by observations of hard X-ray oscillations \citep{makis14} and periodicity in repeating fast radio bursts \citep{lev20,zan20}. Aside from variation in the spindown luminosity, we find that the precessing fits in general predict a slightly faster rotation rate (see Fig. \ref{fig:pdist}) and a slightly lower polar magnetic field strength (see Fig. \ref{fig:bdist}) over their orthogonal counterparts. A faster rotation rate may help to stabilise the post-merger remnant against gravitational collapse \citep{breu18,rezz18}, thereby strengthening the magnetar central-engine model in general since a greater range of pre-merger NS masses may be permitted \citep{giac13,lu15,ciolfi17}.

Since the precession frequency is directly proportional to the spin-misaligned ellipticity, we can, assuming that the stellar deformation arises due to a mixed poloidal-toroidal internal field [though cf. \cite{don15,kru20,kru20b}] and that the internal field strength is of the same order as the external field strength [though cf. \cite{pav02,shab12}], estimate the relative strength of the toroidal sector. We achieve this by introducing the parameter $\Lambda$, defined as the poloidal-to-toroidal magnetic energy ratio, and compare the fitted ellipticities with those coming from some canonical model of magnetic deformations; see equation \eqref{eq:epsilon} and Fig. \ref{fig:lambdas}. We find that the distribution of poloidal-to-toroidal field strengths is bimodal, clustering around energetically-equal (8 candidates; $\Lambda \approx 0.38$) and toroidally-dominated (8 candidates; $\Lambda < 0.1$) partitions, suggesting perhaps that substantial field-line winding via differential rotation occurs in $\sim 50\%$ of newborn NSs.

It is important to note that our model for NS spindown is very simple, precessing or otherwise, and is meant to serve as a proof of concept rather than being representative of a realistic astrophysical environment. For example, we neglect spindown due to non-dipole fields {[a pure $\ell$-pole spins down at a rate $\dot{\Omega}_{\ell} \sim \dot{\Omega}_{\text{dipole}} (R \Omega / c )^{2 \ell - 2}$ \citep{krol91,pet19}, which can be large for millisecond objects]}, fallback accretion {[which can not only increase the stellar mass by up to $\sim 0.1 M_{\odot}$ \citep{ross07} and spin-up the star by a factor $\gtrsim 2$ \citep{hask20}, but also gradually reduce the dipole moment by a factor $\gtrsim 10$ through burial \citep{mel14,li20} and lead to the growth of comparable strength $\ell \lesssim 5$-moments \citep{suvm20}]}, mass ablations from neutrino winds {[reducing the moment of inertia at a rate $\dot{I}_{0} \lesssim 4 \times 10^{41} \text{ g cm}^2 \text{ s}^{-1}$ \citep{dess09} and baryon-polluting the surroundings, reducing the radiative efficiency $\eta$ by $\lesssim 2$ orders of magnitude \citep{xiao19}]}, and gravitational radiation {[with relative luminosity contribution $L_{\text{GW}}/L_{\text{EM}} \sim 0.3 \left( \epsilon_{p,-3}^2 B_{p,15}^{-2} P_{0,-3}^{-2} \right)$ at early times \citep{don15,gao20}]}. 

The efficacy of the last of these effects in particular is directly tied to the value of the stellar ellipticity. Therefore, in general, the precession frequency is not totally independent from the other parameters of the system, since gravitational waves help to spin down the star, and may dominate the profile in certain cases \citep{sarin18}. Including these effects in a future work would be worthwhile, especially to investigate the extent to which gravitational waves associated with SGRB afterglows may be detectable by existing and upcoming ground-based interferometers: for rapidly rotating systems with $\Lambda \ll 0.1$, a signal may be detectable out to $\sim 20$ Mpc by Advanced LIGO and $\sim 450$ Mpc by the Einstein Telescope \citep{sarin18} [see also \cite{dall18}], especially if the star is CFS unstable \citep{don15,kru20,kru20b}. Similarly, the quasi-normal ringdown of a newborn black hole, formed either promptly or after $\sim 10^{2}$ s in the case of supramassive stars, may be detectable out to comparable distances. {On the other hand, if a long ($>1$ s) time delay is observed between the peak gravitational-wave emission and the prompt gamma-ray emission, this would strongly support the `time-reversal' scenario considered by \cite{cio15} [see also \cite{rez15}] where a meta-stable magnetar is born out of a merger but the jet is only launched once it collapses, thereby introducing a lag between the gravitational and electromagnetic emissions \citep{ciolfi18}. In the time-reversal scenario, the plateau duration is associated with a diffusion timescale across the environment rather than the longevity of the NS, though precession-induced modulations in the luminosity may still be visible.}

\section*{Acknowledgements}
This work was supported by the Alexander von Humboldt Foundation and by the DFG research grant 413873357. This work makes use of data supplied by the UK Swift Science Data Centre at the University of Leicester. {We thank the anonymous referee for providing helpful feedback.}
\\
\\
Data availability statement: Observational data used in this paper are publicly available in the cited works. The data generated from computations are reported in the body of the paper, and any additional data will be made available upon reasonable request to the corresponding author.

\appendix

\section{Tests on the reliability of the AIC}

In this Appendix, we present some tests of the AIC as a measure for whether some underlying data is best (information-theoretically speaking) represented by a precessing, oblique rotator or by an orthogonal rotator. 

To achieve this, we simulate mock data with some random errors and then run Monte Carlo simulations (as detailed in Sec. 3). We present here four such fake data sets corresponding to two different ``true'' models but with small and large error cases. Although it is difficult to simulate realistic GRB data since their spectral profiles are rather diverse \citep{ciolfi18}, some reasonable approximations can be made. Specifically, we build a temporal grid which consists of uniform spacings in time but with some stochastic variations drawn from a uniform distribution $U$; in this way, the mock times since BAT trigger have a random quality. Furthermore, since most GRB light curves show significant time-gaps in the data due to several factors (e.g., passage of the South Atlantic Anomaly, high particle background, or electronics), we build a total grid that consists of two sections: 
\begin{itemize}
\item[i)]{$\underset{j \in [0,70]}{\bigcup} \{100 + 20j + U(0,5)\}$, and}
\item[ii)]{$\underset{j \in [0,48]}{\bigcup} \{3000 + 250j + U(10,100)\}$.}
\end{itemize}
As such, the temporal grid consists of the following list of points: $100 + U(0,5), 120 + U(0,5), \ldots, 1500 + U(0,5), 3000 + U(10,100), 3250 + U(10,100),  \ldots, 15000 + U(10,100)$, where we sample from $U(a,b)$ via the RandomReal command in Mathematica. There is thus a `gap' between $t \approx 1500$ s and $t \approx 3000$ s, and we have a lower resolution in the second section, which lasts until $t \approx 1.5 \times 10^{4}$ s. This mock data thus represents a case where observations of the afterglow begin $\approx 100$ s after prompt emission is observed and terminate at $t \approx 1.5 \times 10^{4}$ s. Fluxes are generated at each grid point from some model, again with some random errors of various size. 

\begin{figure*}
\includegraphics[width=\textwidth]{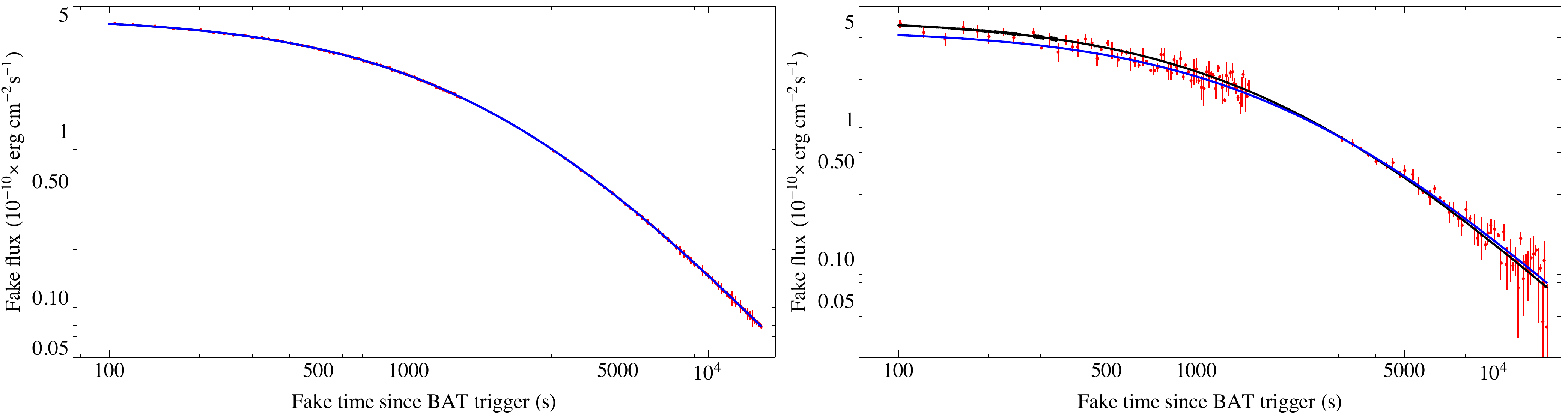} 
\caption{A particular realisation of a mock temporal grid with some associated fluxes, where the underlying data (red points) are drawn from the \emph{orthogonal rotator} distributions with small errors (left panel) and large errors (right panel); see text for details. Over-plotted are the best fits obtained from the Monte Carlo fitting, for the orthogonal rotator (blue curve) and precessing rotator (black curve; invisible for small errors).} \label{fig:fakedorth}
\end{figure*}

To simulate `small' errors and build a model that is almost perfectly described by an orthogonal rotator, fake fluxes are generated at each point $t_{i}$ through $F_{\text{orth}}(t_{i}) + U(-10^{-2},10^{-2})$ on the first section [i)] and $F_{\text{orth}}(t_{i}) + U(-10^{-4},10^{-4})$ on the second section [ii)], where $F_{\text{orth}}(t) = f_{0} \left( 1 + t / \tau \right)^{-2}$ is a normalized functional form for the flux [cf. equation \eqref{eq:spiinlum}]. We take, in CGS units, $f_{0} =  5 \times 10^{-10}$ and $\tau = 2 \times 10^{3}$ for illustration (other values lead to basically the same result). We further introduce some synthetic error bars, drawn from the uniform distribution $U(-10^{-2},10^{-2})$, which are used to weight the Monte Carlo fitting for both the orthogonal and precessing fits. The `larger' error cases are similarly produced but with greater variances: $F_{\text{orth}}(t_{i}) + U(-0.5,0.5)$ on the first section [similarly with error bars from $U(-0.5,0.5)$] and $F_{\text{orth}}(t_{i}) + U(-0.05,0.05)$ on the second section [with error bars from $U(-0.05,0.05)$]. 

Mock data for one particular realisation, together with the results of our fitting code, are shown in Figure \ref{fig:fakedorth}. We see that the fits pull out parameters which are very close to the `true' ones even in cases with relatively large errors (right panel). For the precessing fits, the superfluous parameters are all found to be close to zero (almost exactly zero for the small error case; left panel): the random errors do not entice the fits to conclude that precession is active, despite the oscillatory nature of errors that come about when sampling from the uniform distribution. The AIC numbers, computed from expression \eqref{eq:aicn}, for these fits are shown in Table \ref{tab:fakefits}. We find that the AIC numbers of the orthogonal fits are smaller than their precessing counterparts. Recalling that smaller AIC numbers imply a more reliable fit \citep{ak74,akaike}, we therefore conclude that the orthogonal fits are preferred, as expected. Note that for the small error case, the AIC for the precessing fit is larger by exactly (within machine precision) six, i.e. $\text{AIC}_{p} - \text{AIC}_{\perp} = 6$. As discussed in Sec. 3.1, this is expected from expression \eqref{eq:aicn}, since the precessing model has three additional parameters over the orthogonal rotator, and the fits are identical. Re-initializing the mock data construction (temporal grids, random errors, and/or error bars) and rerunning the Monte Carlo fitter over different data sets consistently leads to essentially the same result. Rare exceptions occur when the errors are so large that little information can be gleaned from the synthetic light curves, and the fits converge to inaccurate values. 

\begin{figure*}
\includegraphics[width=\textwidth]{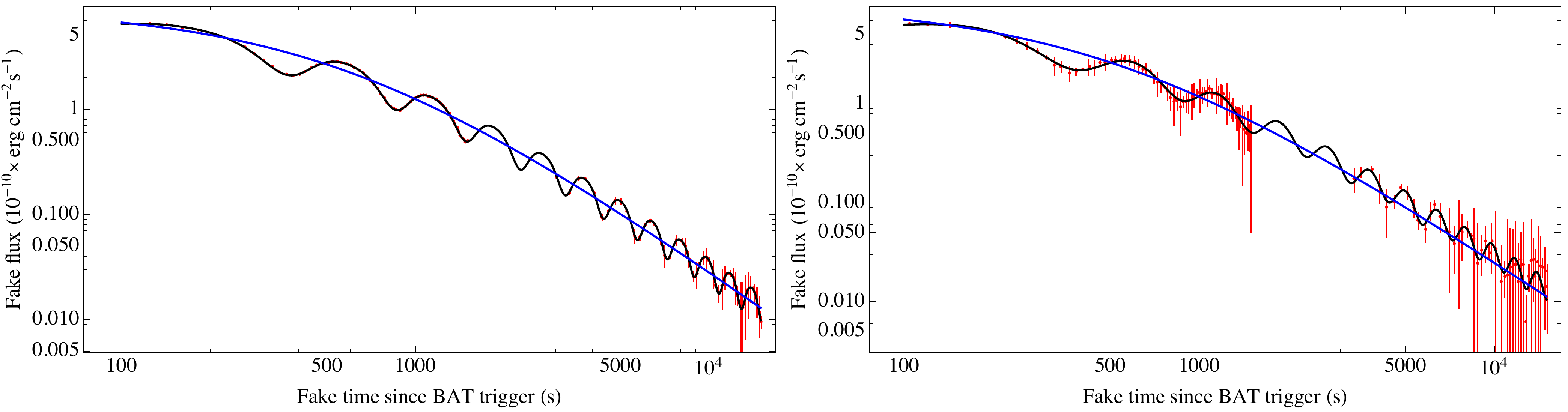} 
\caption{Similar to Fig. \protect\ref{fig:fakedorth}, though with underlying data drawn from distributions appropriate to a \emph{precessing, oblique rotator}.} \label{fig:fakedprec}
\end{figure*}

In the same way as above, we can also test cases where the underlying data corresponds to a precessing rotator. This case is simulated by using expression \eqref{eq:genspinlum}, where the prefactor is replaced by some normalized number $f_{0}$, as in the orthogonal case. We take $\alpha_{0} = 0.7, k= 0.3, \tau = 10^3, \Omega_{p,0} = 0.02$, and $f_{0} = 5 \times 10^{-10}$ for demonstration. The fitting results for one particular realisation are shown in Figure \ref{fig:fakedprec}. As expected, the AIC numbers are lower for the precessing cases (see Tab. \ref{tab:fakefits}). For the small error case, the difference is enormous ($\text{AIC}_{\perp} \gtrsim 10^{4} |\text{AIC}_{p}|$), indicating an obvious preference for precession; the orthogonal model cannot replicate the precessing-source data to any reasonable degree of accuracy. On the other hand, the Monte Carlo simulations converge to the mock parameters to within $\sim 20\%$ (which is of the same order as the variance for the large error case) for the precessing fits.

\begin{table}
\caption{AIC numbers for some particular realisations of Monte Carlo fits to synthetic data. }
\centering
  \begin{tabular}{llccc}
  \hline
Test case &  $\text{AIC}_{\perp}$ & $\text{AIC}_{p}$  \\
\hline
Orthogonal, small errors & 735.6 & 741.6  \\
Orthogonal, large errors & 376.5  & 6097 \\
\hline
Precessing, small errors & $1.89 \times 10^{6}$ & -774.9  \\
Precessing, large errors & 853.3 & -360.9 \\
\hline
\end{tabular}
\label{tab:fakefits}
\end{table}

\bsp \label{lastpage}

\end{document}